# Failure of Mineralized Collagen Microfibrils Using Finite Element Simulation Coupled to Mechanical Quasi-brittle Damage

Abdelwahed BARKAOUI, Awad BETTAMER, Ridha HAMBLI

*PRISME laboratory, EA4229, University of Orleans*
*Polytech' Orléans, 8, Rue Léonard de Vinci 45072 Orléans, France*

**Abstract**

Bone is a multiscale heterogeneous materiel of which principal function is to support the body structure and to resist mechanical loading and fractures. Bone strength does not depend only on the quantity and quality of bone which is characterized by the geometry and the shape of bones but also on the mechanical proprieties of its compounds, which have a significant influence on its deformation and failure. This work aim to use a 3D nano-scale finite element model coupled to the concept of quasi-brittle damage with the behaviour law isotropic elasticity to investigate the fracture behaviour of composite materiel collagen-mineral (mineralized collagen microfibril). Fracture stress-number of cross-links and damping capacity-number of cross-links curves were obtained under tensile loading conditions at different densities of the mineral phase. The obtained results show that number of cross-links as well as the density of mineral has an important influence on the strength of microfibrils which in turn clarify the bone fracture at macro-scale.

## 1. Introduction

A microscopic analysis reveals a complex architecture that can be described as follows. The bone is a composite material: it must imagine hollow cylinders juxtaposed next to each other and sealed by a matrix. The cylinders are called Osteon, the inner bore Haversian canal and the matrix pore system. Further analysis shows that osteons are in fact an assembly of cylindrical strips embedded in each other and each blade is composed of a network of fibers wound helically oriented collagen and inserted into hydroxyapatite crystals. The orientation of collagen fibers may be different between two consecutive slices. These fibers are one set of fibrils. Each fibril is in turn composed of microfibrils. Finally, each micro fibril is a helical arrangement of five tropocollagen molecules. Fig. 1 provides a better understanding of this large complex architecture [1].

The existence of sub-structures in collagen fibrils has been a debate for years. Recent studies suggest the presence of microfibrils in fibrils, experimental works by Hulmes, Orgel, Fratzl, Currey, Aladin and others prove that virtually all collagen-based tissues are organized into hierarchical structures, where the lowest hierarchical level consists of triple helical collagen molecules [2-7] and the multi-scale structure was defined as tropocollagen molecular-triple helical collagen molecules-fibrils-fibers. A longitudinal microfibrillar structure with a width of 4 - 8 nm was visualized in both hydrated [8-10] and dehydrated [11]. Three-dimensional image reconstructions of 36 nm-diameter corneal collagen fibrils also showed a 4 nm repeat in a transverse section, which was related to the microfibrillar structure [12]. Using X-ray diffraction culminating in an electron density map, Orgel et al. [3] suggested the presence of right-handed super twisted microfibrillar structures in collagen fibrils.

Bone quality is prescribed by collagen characteristic including collagen cross-links which are important in reinforcement of bone strength. Biomechanical effects of collagen depend largely on cross-linking [13-16]. The strength and stability during maturation of the microfibrils are achieved by the development of intermolecular cross-links [17-18].

In Summary mineralized collagen microfibril is a composite material containing an inorganic mineral hydroxyapatite reinforced by tropocollagen molecules linked together by cross-links. The mineral phase is almost entirely of crystalline hydroxyapatite $Ca_{10}(PO_4)_6(OH)_2$ impure with some phosphate. This happens, a mature form of needles or leaves, included into the gap and between the tropocollagen molecular. The size of human cortical femur bone hydroxyapatite crystals reported in the literature varies, with values ranging from length, 53.5(22.5-110) nm; width, 28.5(15-60) nm; thickness, 2-10 nm. [19-20].

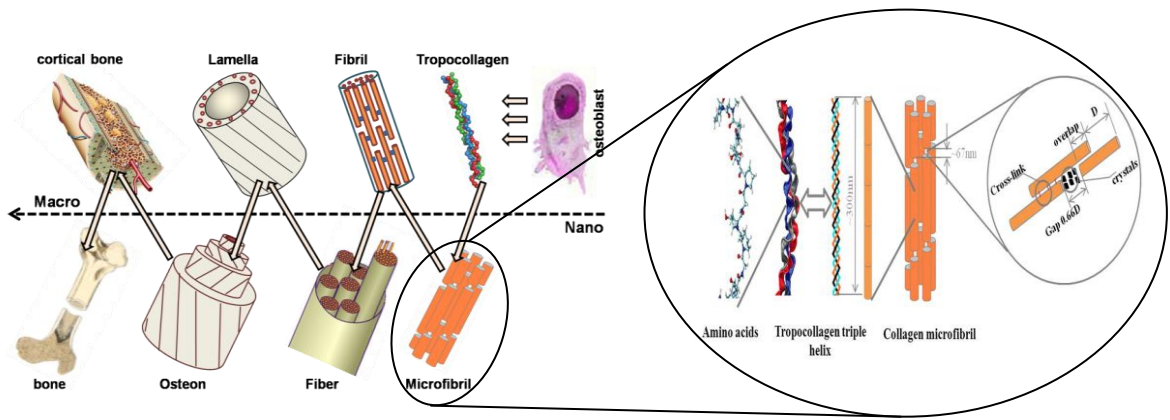

**Figure.1** The multiscale hierarchical structure of cortical bone and schematic illustration of the formation of the microfibril.

The fracture resistance of bone at organ level (whole bone) depends on its multiscale organization which depends mainly on the microfibrils composition and its substructure. A parametric study was performed in order to investigate the influence of varying number of cross-links on the mechanical behavior of the microfibril. A Python/Fortran script programs were developed in order to compute the averaged microfibril failure properties based on optimization procedure using Abaqus code.

## 2. Methods

*Finite element model:*

A three-dimensional finite element model of collagen microfibril with symmetric and periodic boundary conditions is developed here, with an array of 5 tropocollagen molecules cross-linked together using spring elements, the whole structure is embedded into a mineral matrix Fig. 2.

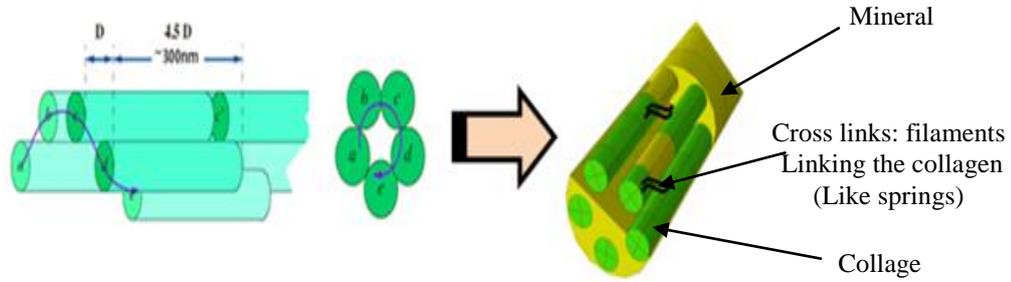

**Figure.2** Real and finite element model of collagen microfibril

*Mechanical Quasi-brittle Damage law:*

To model the progressive development of damage of bone through the decrease in its elastic stiffness, a homogenized measure of damage is introduced, which in the simplest case is represented by a scalar *D*, called the damage variable, which has values in the range (0,1). The stress-strain relation of elasticity based damage mechanics is expressed by [22]:

$$\sigma_{ij} = (1-D)\, C_{ijkl}\, \varepsilon_{kl} \quad \text{with} \quad 0 < D < 1 \qquad (1)$$

Where: $\sigma_{ij}$ the Cauchy stress components, $\varepsilon_{kl}$ the linear strains and $C_{ijkl}$ the components of elasticity tensor.

Due to the brittle nature of the mineral and ductile nature of collagen, a quasi-brittle damage law was retained in the present work to describe the progressive failure of the microfibril. Experiment data showed that the tensile curve exhibits a typical behavior of quasi brittle material.

A quasi brittleThe damage law retained in the present work is Combining Nagaraja's [22] experimental law and the dependence of damage growth to the hydrostatic pressure, a quasi-brittle damage law can be expressed by:

$$\begin{cases} D = 0 & ;\ \varepsilon_{eq} \leq \varepsilon_0 \\ D = (\varepsilon_{eq}/\varepsilon_f) & ;\ \varepsilon_0 < \varepsilon_{eq} < \varepsilon_f \\ D = 1 & ;\ \varepsilon_{eq} \geq \varepsilon_f \end{cases} \qquad (2)$$

Table.1 mechanical properties of collagen and mineral phase:

|  | Young's modulus (GPa) | Poisson's ratio | fracture strain |
| --- | --- | --- | --- |
| Collagen | 2.7 | 0.35 | 0.002 |
| Mineral | 114 | 0.27 | 0.02 |

*Damping capacity:*

This quantity without dimension expresses the ease with which a material can dissipate vibrational energy. Elastically loaded material with a constraint **σ** stores energy per volume unit:

$$U = \int_0^\varepsilon \sigma d\varepsilon = \varepsilon^2/2E \qquad (3)$$

In a cycle of loading and unloading, it dissipates energy:

$$\Delta U = \oint \sigma d\varepsilon \qquad (4)$$

The damping capacity is:

$$\eta = \Delta U / 2\Pi U \qquad (5)$$

## 3. Results and discussion

Studies in vitro [23-24] and in vivo [25-26] have reported that increases in cross-linking are associated with enhancement of some mechanical properties (strength and stiffness) and reductions of others (energy absorption) justified in this study the formula for calculating the damping capacity η= ΔU/2ΠU .These experimental data are limited in their ability to define individual biomechanical effects of altered cross-linking [27]. In the present paper the study of the effect of cross-links is more precise and realistic that's because the proposed finite element model success to combines the collagen, mineral and cross-links furthermore being able to vary the mechanical and geometrical properties of each phase and to visualize their influence on corresponding equivalent properties of the collagen microfibril.

Fig.4 depicts the fracture stress as a function of the cross-link number curve for various fraction volume of mineral. It is found that as the fraction volume of mineral increases, the collagen microfibrils becomes hard and the fracture stress increases. The graph also shows that the number of cross-link has more influence on the increase in the fracture stress. if the number of cross-link increases the bone material will be more stronger, however when N > 20 fracture stress does not depend on the cross-link number and it remains at constant value "plateau value", same observation has been also found by Markus J.Buehler (2008) in his molecular multi-scale study. Where it is found that the yield stress and fracture stress depend on the cross-link density" β" only when β < 25 and it is interpreted as the plateau value can be explained by a change in molecular deformation mechanism from predominantly shear (for β < 25) to molecular fracture (for β >25) [28]. This plateau can also be explained by the fact that whenever the number of cross link increases their stiffness increases up a threshold value which the behaviour of all collagen-cross-links becomes insensitive to the number, this explanation is demonstrated by an analytical calculation.

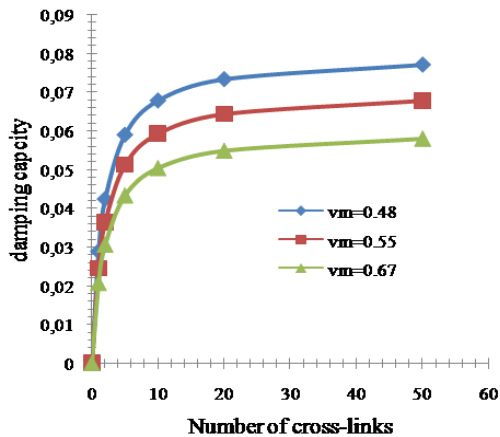
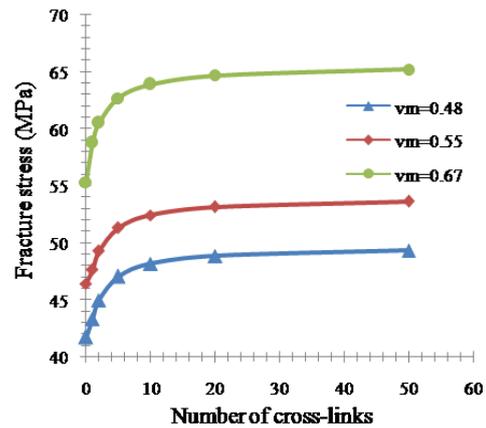

**Figure.3** Damping coefficient of a collagen microfibril as a function of the cross-link number under varying fraction volume of

**Figure.4** Fracture stress of a collagen microfibril as a function of the cross-link number under varying volume fraction of

Fig.3 depicts the damping coefficient-number of cross-link curve for deferent volume fraction of mineral. This graph shows the effect of number of cross-links on the damping capacity. The extent to which collagen is cross-linked is often related to the amount of energy that can be absorbed by the tissue after yield [27],This graph also shows the increase in damping capacity and reductions of the amount of energy absorbed, due to the increased number of cross links.

**4. conclusion**

In this paper we study for the first time the nano-mechanical properties at failure of the microfibril using 3D finite element simulation coupled to mechanical quasi-brittle damage and inverse identification method. Consistent results are found, similar to that found in the literature in macroscopic scale, they show the important role of cross-links that is improving mechanical properties and preventing energy absorption which increases the damping capacity. This work also allows us to understand better this nanoscale and study the upper level scale which is the collagen fibril with the same methods by using the results found in this scale.

**Acknowledgements**

This work has been supported by French National Research Agency (ANR) through TecSan program (Project MoDos, n°ANR-09-TECS-018).